\documentstyle[11pt,aaspp4]{article}

\lefthead{}
\righthead{Probes for Nearby Galaxies}

\begin{document}

\title{Probes for Nearby Galaxies}

\author{David Crampton\altaffilmark{1,2} and G. Gussie\altaffilmark{2,4}}
\affil{Dominion Astrophysical Observatory, National Research Council of
Canada,\\ Victoria, B.C. V8X 4M6, Canada} 

\author{A.P. Cowley\altaffilmark{1,2,3} and P.C.
Schmidtke\altaffilmark{1,3} } \affil{Department of Physics \& Astronomy,
Arizona State University, Tempe, AZ, 85287-1504} 

\altaffiltext{1} {Visiting Astronomers, Cerro Tololo Inter-American
Observatory, National Optical Astronomy Observatories, which is
operated by the Association of Universities for Research in Astronomy,
Inc., under contract with the National Science Foundation}
\altaffiltext{2} {Visiting Astronomers, Canada-France-Hawaii Telescope,
operated by the National Research Council of Canada, the Centre de la
Recherche Scientifique de France, and the University of Hawaii}
\altaffiltext{3} {Guest Observer, Multiple Mirror Telescope
Observatory, operated jointly by the Smithsonian Observatory and the
University of Arizona}
\altaffiltext{4} {Student assistant, 1985. Present address: Dept.\ of
Physics, U.\ of Tasmania, Hobart, Australia} 

\begin{abstract}

Data are given for sixteen extragalactic objects (predominantly AGN)
behind the Magellanic Clouds and for 146 quasar candidates behind the
nearby galaxies NGC 45, 185, 253, 2366, 2403 and 6822, IC 1613, M31 and
M33.  The Magellanic Cloud objects were identified by their X-ray
emission, and precise optical and X-ray positions and optical
photometry and spectra are presented for all of these.  The quasar
candidates surrounding the other nearby galaxies were identified
through a CFHT slitless spectral survey. Although redshifts for only
eight of these candidates have been obtained, previous observations
indicate that the majority are likely to be quasars. A subsample of 49
of the brighter objects could confidently be used, in addition to the
Magellanic Cloud sources, as probes of the gas in nearby galaxies for
rotation curve studies, for studies of their halos, for comparison with
higher redshift QSO absorption lines, or as references for proper
motion studies.

\end{abstract}


\section{INTRODUCTION}

The UV spectroscopic capabilities of HST allow nearby galaxies to be
probed via absorption lines in the same way that QSO absorption line
systems probe galaxies at high redshift.  Although recent studies have
shown that Mg II absorption lines at z $<$ 1 appear to be associated with
luminous, massive galaxies (Steidel, Dickinson \& Persson 1994), there is
still a debate about the nature of the absorbing material.  For example,
it is unclear whether the absorption occurs in an extended disk, halo or
even satellite galaxies.  Studying absorption lines in AGN and QSOs behind
nearby galaxies allow these galaxies to be probed along multiple
sightlines, thereby yielding information on how the absorption correlates
with various parameters of the foreground galaxy.  Accurate measurements
of velocities of the interstellar material can also be used to extend
rotation curves to larger galactocentric distances, thereby providing
improved estimates of the mass distributions.  Furthermore, the background
quasars can be used for precise proper motion studies of the foreground
objects.  Unfortunately, however, most surveys for quasars have avoided
directions towards nearby galaxies.  Monk et al.\ (1986) give a list of
the brightest (m $<$ 17.5) quasars located within $\sim$200 kpc of nearby
galaxies.  In this paper we report results from an X-ray survey of
the Magellanic Clouds and an optical slitless-spectra survey of nearby
northern hemisphere galaxies.

For several years we have been undertaking a census of the X-ray sources
in the Magellanic Clouds, first with data from {\it Einstein} Observatory
(e.g., Cowley et al.\ 1984) and, more recently, with $ROSAT$ (Schmidtke et
al.\ 1994, hereafter Paper I; Cowley et al.\ 1997, hereafter Paper II). 
Several background galaxies, AGN, and QSOs were detected in addition to
X-ray bright objects within the Clouds.  Some of these were reported in
Papers I and II, but the identifications and redshifts of most have only
recently been determined.  In this paper we bring together the data for
all of these objects.  We give photometry and redshifts for sixteen X-ray
selected extragalactic objects, five in the SMC field and eleven in the
vicinity of the LMC.  We also present finding charts for those that are
not already in the literature. 

In the course of another project, searches for quasar candidates were
carried out on a series of CFHT grens plates of nearby galaxies.  These
plates are ideally suited for the detection of QSO candidates (e.g.,
Crampton, Schade \& Cowley 1985). Although we have not been able to
confirm the identifications or measure the redshifts for most of these
candidates, our previous high success rate in identifying QSOs
(Crampton, Cowley \& Hartwick 1987) demonstrates that most of them can
confidently be expected to be AGN. For this reason we present these
data now since, apart from the faintest low quality identifications,
most will be quasars and hence useful as probes regardless of their
redshift. However, a low S/N spectrum confirming the QSO identification
is recommended before investing large amounts of telescope time for,
say, absorption line studies.

\section{X-RAY SELECTED AGN BEHIND THE MAGELLANIC CLOUDS}

Nearly 200 point X-ray sources in the direction of the Magellanic Clouds
were detected by {\it Einstein} Observatory (Long, Helfand, Grabelsky
1981; Wang et al.\ 1991; Seward \& Mitchell 1981, Wang \& Wu 1992; Cowley
et al.\ 1984).  Details of recent $ROSAT$ X-ray observations of many of
these Magellanic Cloud sources are given in Papers I and II. 
Identifications of the optical counterparts have been carried out through
extensive photometric and spectroscopic observations at CTIO.  This
program reveals that many of these sources are associated with foreground
stars or background extragalactic objects (e.g. Papers I and II).  The
optical data which we have obtained are described below. 

Our $ROSAT$-HRI survey of the Magellanic Clouds was aimed at obtaining
improved positions of $Einstein$ sources which had not been
already identified to enable detection of new optical
counterparts.  Therefore, our material does not cover the entire fields
of the LMC and SMC and thus does not comprise a complete sample of AGN
behind these galaxies.  The objects which are included in our X-ray
sample are listed in Table 1 together with their $ROSAT$ positions and
count rates. We have carefully investigated the positional accuracy
delivered by the $ROSAT$ detectors (see discussion in Papers I and II)
and the positions reported here are accurate to about
$\pm5^{\prime\prime}$, making the search for optical counterparts
relatively straightforward.  Figures 1 and 2 show the distribution of
X-ray bright extragalactic objects we have found near these two
galaxies.  Finding charts are given in Figures 3 and 4 for those which
are not already in the literature.  The X-ray positions are indicated
by a `{\bf$+$}' and the optical counterpart is marked with a dash.
Three of the X-ray sources are spatially extended, and for those we
have overlaid their X-ray contours on the optical finding charts.  Each
of these appears to be associated with a cluster of galaxies.

\subsection{Photometry}

All the photometric data were obtained from CCD photometry carried out at
CTIO during observing runs between 1992 and 1996 using the 0.9-m
telescope.  The $V$ magnitudes presented in Table 1 are based on aperture
photometry, calibrated using observations of Landolt (1992) standard
stars.  The accuracy of the magnitudes is about $\pm0.02$ mag.  A few
special cases are mentioned below where the object was extended or
complicated in some way.  The astrometry of each CCD frame used for the
finding charts has been tied to the coordinate system of the {\it HST
Guide Star Catalogue} (Lasker et al.\ 1990) by measuring the positions of
$\sim$6--8 stars on the digitized $GSC$. 

\subsection{Spectroscopy}

Our highest resolution AGN spectra were obtained with the CTIO 4-m
telescope in November 1996 with the KPGL1 grating and Loral 3K detector.
These spectra cover the wavelength range 3700--6700\AA\ and have a
resolution of $\sim$1.0\AA\ per pixel.  With a 1\farcs5 slit,
corresponding to three pixels, the spectral resolution is 3\AA.  The
spectra of the optical counterparts of CAL 21 and RX J0532.0$-$6920 were
taken with the CTIO ARGUS fiber system in December 1995.  These spectra
cover the range 3650-5800\AA\ with a resolution of 1.8\AA\ per pixel.  The
spectrum of RX J0547.8-6745 has a wavelength range of 3720--5850\AA\ and
also has a resolution of 1.8\AA\ per pixel.  One-dimensional spectra were
extracted and processed following standard techniques with {\sc IRAF} to
yield wavelength-calibrated spectra.  The spectra are shown in Figure 5,
shifted to restframe wavelengths according to their redshifts (Table 1). 

Some of the earlier spectra were taken with the SIT vidicon detector on
the CTIO 4-m RC spectrograph.  These are not shown in Figure 5, but the
objects observed are included in Table 1. 

\subsection{Individual Sources} 

Some of the newly identified individual sources deserve comment beyond 
just listing them in Table 1.  Of the ones previously published (see
references in Table 1), we point out that RX J0534.8$-$6739 was only
listed as a ``note added in proof" by Cowley et al.\ (1997) so it would be
very easy to overlook it in that paper.  The AGN is ``star" 2 in the
finding chart for this source in Paper II.

\noindent
RX J0005.3$-$7427 (SMC 1): ~This X-ray point source falls nearly on the
galaxy shown in the finding chart in Fig.\ 3.  Its spectrum, shown in
Fig.\ 5, indicates the object is a Seyfert I galaxy with a redshift of z
$=$ 0.1316.  The optical image shows the galaxy is just resolved. 

\vskip 5pt 

\noindent
RX J0033.3$-$6915 (SMC 70): ~The X-ray contours for this source are very
extended, as shown in Fig.\ 3.  The X-ray position falls on a large cD
galaxy in the center of the cluster Abell 2789.  Thus, the X-rays appear to
result from hot gas in the cluster.  The redshift of the cD galaxy
is z $=$ 0.0975. 

\vskip 5pt 

\noindent
RX J0119.5$-$7301 (SMC 66): ~The extended X-ray contours suggest that this
source also is arises from hot gas, but this is a previously unknown cluster. 
Many galaxies are visible in the field (see finding chart in Fig.\ 3).  A
spectrum of the bright cD galaxy to the west of the central X-ray contour
shows it to have a redshift of z $=$ 0.0658.  The optical image of this
galaxy has three parts, consisting of a foreground star and two
non-stellar condensations.  The magnitude given in Table 1 refers to the
brighter (south-western) of the two non-stellar parts. 

\vskip 5pt

\noindent
RX J0135.4$-$7048: ~ This weak, extended source also appears to be
associated with an unknown cluster of galaxies (visible on our original
image, but not easily seen in Figure 3).  The center of the extended
X-ray contours is not coincident with any bright galaxy, but we have
observed the nearest (bright) one, which is south-east of the third
X-ray contour (as marked on the finding chart in Fig.\ 3).  Its
spectrum gives a redshift of z $=$ 0.0647. The identification of this
X-ray source with the cluster, and whether the galaxy we observed is a
member, should be verified by obtaining redshifts of some of the
fainter galaxies.

\vskip 5pt 

\noindent
RX J0136.4$-$7105 (SMC 68), RX J0454.2$-$6643, and RX J0550.5$-$7110: ~
These three sources are all associated with AGN, as shown by their
redshifts in Table 1 and their spectra in Fig.\ 5. 

\vskip 5pt

\noindent
RX J0534.0$-$7145: ~ This X-ray point source appears to be located in
or near to the nucleus of the very large optical galaxy, Up
053448$-$7147.3, which is listed as an S0(r).  The $V$ magnitude was
measured in an 80$^{\prime\prime}$ aperture.  The optical spectrum
shows narrow [O II] and [O III] emission but otherwise is relatively
normal. It thus appears to be one of the narrow-emission-line galaxies
that comprise about $\sim$15\% of the extragalactic X-ray galaxy
population (e.g., Griffiths et al. 1996).  The large extent of the optical
disk can be seen in Fig.\ 4.

\vskip 5pt
\noindent
RX J0547.8$-$6745: ~This point X-ray source is identified with an AGN with
redshift z $=$ 0.3905.  It was also recently found to be a compact radio
source, MDM 100 (Marx et al.\ 1997).

\section{QSOS AND QSO CANDIDATES AROUND NEARBY GALAXIES}

Quasars with z $<$ 3.4 and $B <$ 21 can be easily recognized from CFHT
blue grens images.  The blue grens, a grating-prism-lens combination
designed by E.H. Richardson, produces spectra with a dispersion of 945\AA\
mm$^{-1}$ and a wavelength range of 3500-5300\AA\ when recorded on IIIaJ
emulsion.  The grens plates cover a 55\arcmin $\times$ 55\arcmin ~field,
although some parts of this field may be vignetted by the guide probe.
Grens exposures were obtained of eight fields centered on northern nearby galaxies and two ``halo" M31 fields, centered on field C29 of
Sargent et al.\ (1977) and a field to the SW, primarily to study objects
in the galaxies themselves.  In some cases, observations with different
grens orientations were taken to alleviate problems of overlapping images.
A list of the plates is given in Table 2.  Unfortunately, the seeing was
not very good ($> 1\arcsec$) during most of these observations. 

As in previous surveys, four visual searches for objects with a UV excess
and/or emission lines were made of each plate by at least two of the
authors (in this case, by DC, GG and APC).  Objects satisfying these
criteria but which were likely to be associated with the nearby galaxy
were ignored in this survey.  Subsequently, PDS scans were made of the
spectra of all candidates with a 50 micron square aperture.  These spectra
were converted to intensity versus wavelength with a software package
written by Graham Hill.  As in previous quasar surveys (e.g., Crampton,
Schade \& Cowley 1985), the candidates were assigned a class based on
these tracings and a final visual inspection of the original image.  Class
1 candidates are certain quasars with strong emission lines, class 4
objects show UV excess with no definite spectral features, and classes 2
and 3 are intermediate between these extremes.  Recognizable white dwarfs
are included as class 5 since they might otherwise be selected as quasar
candidates on the basis of their colors.  Emission-line galaxies with no
significant spatial extension or ``extragalactic H II regions" are
assigned class 6.  Spectroscopic follow-up observations with the MMT of
candidates with m $<$ 20.5 indicate that, on average, 100\% of class 1
objects are quasars, 93\% of class 2, 76\% of class 3 and 42\% of class 4
candidates are quasars.  Further details of the observational and
identification procedures are given by Crampton, Schade \& Cowley (1985). 
Due to the low galactic latitude of many of these fields, and to internal
absorption in the galaxies themselves, fewer candidates than typical were
identified in these fields.

Positions and magnitudes of all candidates were measured from glass copies
of the Palomar Sky Survey O plates using the method and software described
by Stetson (1979).  Scans of 100$\times$100 10$\micron$ square pixel boxes
were made of each candidate with the PDS and the positions were related to
nearby SAO stars.  Subsequently, the new {\sc SKYCAT} software was used in
conjunction with the digitized Palomar Observatory Sky Survey to
double-check the coordinates and charts and, in some cases, to correct for
errors.  The resulting positions are accurate to $\sim2\arcsec$ and
magnitudes to $\sim$0.3 mag. 

A list of all candidates is given in Table 3.  The first column gives the
candidate name derived from truncated 2000 coordinates in the form
HHMM.M+DDMM.  An internal identification symbol is given in the second
column, followed by the 2000 coordinates, the magnitude as estimated from
the POSS plates, and the class or certainty of the identification.  In the
notes column we give: (1) the estimated wavelengths of any emission
features visible on the grens spectra, listed in order of decreasing
intensity (2) redshifts, listed to one digit accuracy if they were
estimated from the grens observations, (3) other comments or remarks (for
explanation of the abbreviations and any measured redshifts, see the Notes
to the table).  Rather than give a table listing projected distances from
the centers of the nearby galaxies, the locations of the candidates
(marked with their ID as given in Table 3) are shown in Figures 6 -- 15
so that their distances relative to the optical extent of the galaxies
are obvious. 

Spectra of nine of the candidates in NGC 2366 and NGC 2403 were obtained
in rather poor weather conditions with the MMT in 1987 February with the
photon-counting spectrograph.  The spectra cover the 3000--8000\AA\ region
with a resolution of $\sim$7\AA.  Identified features and redshifts for
these candidates are given in the notes to Table 3.

\section{SUMMARY}

Sixteen galaxies, clusters, and AGN behind the Large and Small Magellanic
Clouds have been identified through their X-ray emission.  Redshifts of
most of these indicate that they are relatively nearby, with only three
having z $>$ 0.3.  Quasar candidates in ten fields in the direction of
nearby galaxies have been identified on the basis
of their colors and emission lines.  Of these, twenty-two have magnitudes
brighter than m$=$19 and thus are excellent targets for high spectral
resolution studies with HST or 10-m class telescopes. 
Forty-nine candidates with magnitudes brighter than m$=$20.5 and classes 1
-- 3 have an extremely high probability of being quasars, based on similar
surveys. 

\acknowledgments

We thank Dr.\ Martha Hazen of the Harvard College Observatory who located
the photographs of the Magellanic Clouds and kindly sent them to us, and
Dr.\ Sidney van den Bergh for permission to use his plate of M 33.  We
also thank Y.\ Yuan for her careful checking of the material in Table 3
and for assistance with the diagrams.  The excellent new {\sc SKYCAT} tool
developed jointly by ESO and the CADC was extremely useful in verifying
and confirming positions and magnitudes of all our objects.  We thank D.\
Durand for his support in its use and help in installing additional
catalogs.  A.P.C. and P.C.S. gratefully acknowledge support from NSF for
this work.

\clearpage

\begin{figure}
\caption{The Large Magellanic Cloud, showing the distribution of X-ray
galaxies and AGN discussed in this paper.  The photograph was taken with
the 8-inch Bache photographic doublet at the Boyden Station of the Harvard
College Observatory in Arequipa, Peru on 1897 Nov.\ 25.  This 6-hour and
5-minute exposure was taken using a blue-sensitive emulsion.  (Harvard
plate no.\ B20849, reproduced with the permission of Harvard College
Observatory)} 
\end{figure}

\begin{figure}
\caption{The Small Magellanic Cloud, showing the distribution of X-ray
galaxies and AGN discussed in this paper.  The photograph was taken with
the 10-inch Metcalf photographic triplet at the Boyden Station of the
Harvard College Observatory in Bloemfontein, South Africa on 1950 Oct.\ 4.
This 45-minute exposure was taken using a blue sensitive emulsion.
(Harvard plate no.\ MF39311, reproduced with the permission of Harvard
College Observatory)} 
\end{figure}

\begin{figure}
\caption{Finding charts for the Magellanic Cloud X-ray sources RX
J0005.3$-$7427 (SMC 1), RX J0033.3$-$6915 (SMC 70), RX J0119.5$-$7301 (SMC
66), RX J0135.4$-$7048, RX J0136.4$-$7105 (SMC 68), and RX J0454.2$-$6643.
Note that the finding charts have different scales.  For the extended
X-ray sources (SMC 70, SMC 66, and RX J0135.4$-$7048) the X-ray contours
have been overlaid on the optical images.  These three are associated with
clusters of galaxies.  For the point X-ray sources, the X-ray position is
shown by the $+$ and the optical identification is marked by a dash.} 
\end{figure}

\begin{figure}
\caption{Finding charts for RX J0534.0$-$7145 and RX J0550.5$-$7110.  The
X-ray contours are shown for RX J0534.0$-$7145, although it is a point
source.  The large, central galaxy is Up 053448$-$7147.3, S0(r).  For RX
J0550.5$-$7110 the X-ray position is shown by a $+$ and the associated AGN
is marked by a dash.  Note that the two charts have different scales.} 
\end{figure}

\begin{figure}
\caption{Spectra of galaxies and AGN associated with X-ray sources behind
the Magellanic Clouds.  The spectra have been `de-redshifted' and gaussian
smoothed for display purposes.  They were then normalized to a peak
continuum intensity of one and arbitrarily shifted so that they could be
presented in order of increasing redshift from bottom to top.  Rest
wavelengths of some prominent features are indicated.  The sources are:
(1) RX J0119.5$-$7301, the central galaxy in a cluster, (2) RX
J0534.0$-$7145, (3) RX J0033.3$-$6915, (4) RX J0135.4$-$7048, (5)
0534.8$-$6739, Sy I galaxy, (6) RX J0005.3$-$7427, Sy I galaxy, (7) RX
J0517.3$-$7044, (8) RX J0532.0$-$6920, (9) RX J0454.2$-$6643, (10) RX
J0531.5$-$7130, (11) RX J0547.8$-$6745, (12) RX J0550.5$-$7110, and (13)
RX J0136.4$-$7105.} 
\end{figure}

\begin{figure}
\caption{NGC 45, showing the distribution of quasar candidates and white
dwarfs.  The image is taken from the Sky Survey print.  North is to the top
and East to the left, and the bar indicates 5\arcmin.} 
\end{figure}

\begin{figure}
\caption{M 31 -- Field E, showing the distribution of quasar candidates
and white dwarfs.  The image is taken from the Sky Survey print.  North is
to the top and East to the left, and the bar indicates 5\arcmin.} 
\end{figure}

\begin{figure}
\caption{M 31 -- Field C29 (Sargent et al. 1977), 
showing the distribution of quasar candidates
and white dwarfs.  The image is taken from the Sky Survey print.  North is
to the top and East to the left, and the bar indicates 5\arcmin.} 
\end{figure}

\begin{figure}
\caption{NGC 185, showing the distribution of quasar candidates and white
dwarfs.  The image is taken from the Sky Survey print.  North is to the top
and East to the left, and the bar indicates 5\arcmin.} 
\end{figure}

\begin{figure}
\caption{NGC 253, showing the distribution of quasar candidates and white
dwarfs.  The image is taken from the Sky Survey print.  North is to the top
and East to the left, and the bar indicates 5\arcmin.} 
\end{figure}

\begin{figure}
\caption{IC 1613, showing the distribution of quasar candidates and white
dwarfs.  The image is taken from the Sky Survey print.  North is to the top
and East to the left, and the bar indicates 5\arcmin.} 
\end{figure}

\begin{figure}
\caption{M 33, showing the distribution of quasar
candidates and white dwarfs from all three fields listed in Tables 2 and
3.  The image is reproduced from a Palomar 5-m
103aO plate taken by S.\ van den Bergh.  North is to the top and East to
the left, and the bar indicates 5\arcmin.} 
\end{figure}

\begin{figure}
\caption{NGC 2366, showing the distribution of quasar candidates and white
dwarfs.  The image is taken from the Sky Survey print.  North is to the top
and East to the left, and the bar indicates 5\arcmin.} 
\end{figure}

\begin{figure}
\caption{NGC 2403, showing the distribution of quasar candidates and white
dwarfs.  The image is taken from the Sky Survey print.  North is to the top
and East to the left, and the bar indicates 5\arcmin.} 
\end{figure}

\begin{figure}
\caption{NGC 6822, showing the distribution of quasar candidates and white
dwarfs.  The image is taken from the Sky Survey print.  North is to the top
and East to the left, and the bar indicates 5\arcmin.} 
\end{figure}


\clearpage

\begin{references}

\reference{} Chaffee, F.H. et al. 1991, AJ, 102, 461

\reference{} Clowes, R.G. \& Savage, A. 1983, MN, 204, 365

\reference{cow84} Cowley, A.P., Crampton, D., Hutchings, J.B.,
Helfand, D.J., Hamilton, T.T., Thorstensen, J.R., \& Charles, P.A. 1984,
\apj, 286, 196

\reference{} Cowley, A.P., Schmidtke, P.C., McGrath, T.K., Ponder, A.L., 
Fertig, M.R., Hutchings, J.B., \& Crampton, D. 1997, \pasp, 109, 21 (Paper 
II)

\reference{} Crampton, D., Cowley, A.P., \& Hartwick, F.D.A. 1987, \apj,
314, 129 

\reference{} Crampton, D., Schade, D., \& Cowley, A.P. 1985, \aj, 90, 987

\reference{} Cristiani, S. \& Tarenghi, M. 1984, A\&A, 132, 351

\reference{} Garilli, B., Bottini, D., Maccagni, D., Vettolani, G., \&
Maccacaro, T. 1992, AJ, 104 

\reference{} Griffiths, R.E., Della Ceca, R., Georgantopoulos, I., 
Boyle, B.J., Stewart, G.C., Shanks, T. \& Fruscione, A. 1996, MNRAS, 281, 71

\reference{} Landolt, A.U. 1992, \aj, 104, 340

\reference{} Lasker, B.M., Sturch, C.R., McLean, B.J., Russell, J.L, 
Jenkner, H., and Shara, M.M. 1990, \aj, 99, 2019

\reference{lon81} Long, K.S., Helfand, D.J., \& Grabelsky, D.A. 1981,
\apj, 248, 925

\reference{} Marx, M., Dickey, J.M., \& Mebold, U. 1997, A\&A, in press

\reference{} Monk, A.S., Penston, M.V., Pettini, M., \& Blades, J.C. 1986,
MN, 222, 787 

\reference{} Sargent, W.L.W., Kowal, C.T., Hartwick, F.D.A., 
\& van den Bergh, S. 1977, \aj, 82, 947

\reference{} Schmidtke, P.C., Cowley, A.P., Frattare, L.M., McGrath, T.K.,
Hutchings, J.B., \& Crampton, D. 1994, \pasp, 106, 843 (Paper I)
 
\reference{} Seward, F.D. \& Mitchell, M. 1981, \apj, 243, 736

\reference{} Steidel, C.C., Dickinson, M., \& Persson, E. 1994, \apjl,
437, L75 

\reference{} Stetson, P.B. 1979, \aj, 84, 1056

\reference{} Tytler, D. \& Fan, X.-M. 1992, \apjs, 79, 1

\reference{wan91} Wang, Q., Hamilton, T., Helfand, D.J., \& Wu, X. 1991,
\apj, 374, 475

\reference{wan92} Wang, Q. \& Wu, X. 1992, \apjs, 78, 391


\end{references}
\end{document}